\documentclass[]{raa}

\usepackage{graphicx,times}
\usepackage{natbib}
\usepackage{amssymb,amsmath}
\bibpunct{(}{)}{;}{a}{}{,}

\usepackage[pagebackref=true]{hyperref}

\begin{document}

   \title{Insight-HXMT observations of the 2023 outburst in Aql X$-$1}

 \volnopage{ {\bf 20XX} Vol.\ {\bf X} No. {\bf XX}, 000--000}
   \setcounter{page}{1}

   \author{Zhe Yan
   \inst{1,2}, Guobao Zhang\inst{1,2}, Yu-Peng Chen\inst{3},  Mariano Méndez\inst{4}, Jirong Mao\inst{1,2}, Ming Lyu\inst{5,6}, Shu Zhang
      \inst{3}, Pei Jin\inst{4}
   }

   \institute{ Yunnan Observatories, Chinese Academy of Sciences, Kunming 650216, People's Republic of China; {\it zhangguobao@ynao.ac.cn}\\
        \and
             University of Chinese Academy of Science, Beijing 100049, People's Republic of China\\
	\and
Key Laboratory of Particle Astrophysics, Institute of High Energy Physics, Chinese Academy of Sciences, Beijing 100049, China\\
\and 
Kapteyn Astronomical Institute, University of Groningen, P.O. BOX 800, 9700 AV Groningen, The Netherland\\
\and 
Department of Physics, Xiangtan University, Xiangtan, Hunan 411105, People’s Republic of China\\
\and 
Key Laboratory of Stars and Interstellar Medium, Xiangtan University, Xiangtan, Hunan 411105, People’s Republic of China\\
\vs \no
   {\small Received 20XX Month Day; accepted 20XX Month Day}
}

\abstract{
We conducted an analysis of the continuum during the onset and initial decline phases of the 2023 outburst in transient neutron star low-mass X-ray binary Aql X$-$1 using broadband observations from the \textit{Insight-Hard X-ray Modulation Telescope (Insight-HXMT)} instrument. 
To determine the most appropriate model for the continuum of this outburst, we employed three models to explore the evolution of the spectral component. 
These observations revealed that the source transitions from the hard state to the soft state. 
The disk-corona and sphere-corona models both adequately described the spectra of the hard state, while the double blackbody model became preferable after the hard X-ray emission ($>$25 keV) disappeared during the state transition. 
In the soft state, the total emission is dominated by changes in the disk and other blackbody components. 
The combination of the sphere-corona model and the double blackbody model is the most suitable model for this outburst. 
The results suggest that as the source transitioned into the soft state, the emission from the boundary layer was enhanced, and a hot spot occurred. 
Notably, we identified two type-I X-ray bursts, one of which exhibited a significant hard X-ray deficit (significance $\sim$ 4.82 $\sigma$), which indicates that \textit{Insight-HXMT} has the capability to capture the evolution of the corona in a single burst.
}

   \authorrunning{Zhe Yan et al. }            
   \titlerunning{The 2023 outburst in Aql X$-$1}  
   \maketitle

\section{Introduction}           
\label{sect:intro}

   Neutron-star Low-Mass X-ray Binaries (NS-LMXBs) consist a neutron star (NS) and a normal companion star with a mass less than 1 M$_{\odot}$.
   The neutron star forms an accretion disk by accreting material from its companion star, which falls to the surface. 
   These systems can be divided into two categories \citep{2016ApJS..222...15T}: One is the persistent sources, which have persistent radiation, such as 4U 0513$-$40 \citep{1975ApJ...199L..93C}, 4U 1720$-$429 \citep{2005ApJ...621..393W}, 4U 1728$-$34 \citep{2008ApJS..179..360G}, and 4U 1636$-$53 \citep{2023ApJ...944..214W}. 
   The other one is the transient sources, which emits brightening X-ray outbursts irregularly after the quiescent state, lasting from a few days to a few months, such as Aquila X$-$1 \citep[Aql X$-$1, ][]{2012PASJ...64...72S}, XTE J1810-189 \citep{2023MNRAS.526.1154M}, 4U 1608$-$52 \citep{2023JHEAp..40...76C}, and 4U 1730–22 \citep{2023ApJ...942L..12C}.
   
   The occurrence of the outburst is generally interpreted by the disk-instability model \citep{2001NewAR..45..449L}.
   In NS-LMXBs, sudden increases in X-ray intensity have also been observed, characterized by a rapid rise over a few seconds and an exponential decrease over tens to hundreds of seconds.
   These events are called as the thermonuclear (Type-I) X-ray bursts, which are generally believed to be caused by the unstable thermonuclear burning of the accreted hydrogen and/or helium \citep{2008ApJS..179..360G}.
   
   During an outburst, transient NS-LMXBs typically transition from the low/hard state to the high/soft state, then return to the the low/hard state before ultimately entering the quiescent state \citep{1994ApJS...92..511V,1998A&ARv...8..279C,2006ARA&A..44...49R}.
   In the hard state, the spectrum is dominated by the Comptonized component.
   The Compton component is original from the corona that is composed of a hot electron plasma.
   This component up-scatters the soft seed photons from the disk or the NS surface, thereby contributing to the overall spectrum.
   In contrast, during the soft state, the spectrum is dominated by the thermal component.
   The contribution from the accretion disk is widely considered to constitute this thermal component \citep{2007A&ARv..15....1D,2010LNP...794...17G}.
   Given the presence of the solid surface of the NS, the emission of the surface and the boundary layer (BL) may impact the spectrum \citep{2003MNRAS.342.1041D}.
   The BL is formed when the material from the accretion disk falls onto the NS surface and decelerates \citep{1999AstL...25..269I,2008MNRAS.386.1038B}.
   For exploring potential components in the spectrum of the outburst, in addition to the disk and the BL, there may also be the hot spot as a contributing component of the thermal emission, which may originate from restriction of the burning to a small region of the NS surface \citep{2012ARA&A..50..609W}.

   In an effort to elucidate the behavior of the continuum spectrum in transient NS-LMXBs, a multitude of models have been employed in previous research. 
   Among these, two types of model have emerged as the most prevalent.
   Both types incorporate a thermal component and a Compton component; however, the distinction between them lies in the spatial configuration of the Compton component (the corona).
   The first one that can be simplified as the disk-corona model, assumes that the Comptonized seed photons originate from the accretion disk \citep{1988ApJ...324..363W}.
   The second one, simplified as the sphere-corona model, assumes the neutron star itself or the BL contribute the Comptonized seed photons \citep{1989PASJ...41...97M}.
   
   Previous studies have employed these models to elucidate the evolutionary dynamics of the accretion disk, BL, and corona, yielding a series of findings about accretion geometry.
   The disk-corona model has successfully described the spectra of the outbursts in several sources, such as XTE J1710$-$281, 4U 1608$-$52 and Aql X$-$1.
   In the hard state of XTE J1710$-$281, the accretion disk, with an inner disk temperature of $\sim$ 0.3 keV and a radius of $\sim$ 70 km, is Comptonized by the corona with a electron temperature of $\sim$ 5 keV.
   The flux of the disk accounts for $\sim$ 22 percent of the total flux \citep{2020MNRAS.496..197S}.
   In the soft state of 4U 1608$-$52, the temperature of the inner disk becomes $\sim$ 1 keV and the radius becomes $\sim$ 25 km.
   The fraction of the disk flux to the total flux is $\sim$ 61 percent \citep{2024ApJ...971..154B}.
   As Aql X$-$1 transitions from the hard state to the soft state (the outbursts in 2019 and 2020), the temperature of the inner disk increases from $\sim$ 0.6 keV to $\sim$ 0.9 keV, and the inner disk radius increases from $\sim$ 14 km to $\sim$ 20 km  \citep[the covering fraction is fixed at 0.5, ][]{2024MNRAS.532.3961P}.
   Within the disk-corona model, as sources transition from the hard state to the soft state, the temperature and radius of the inner disk increase. 
   The changes are reversed during the opposite transition.
   
   The sphere-corona model has been used effectively in some sources, including SAX J1748.9$-$2021, Aql X$-$1, EXO 0748$-$676 and XTE J1701$-$462.
   In the hard state of SAX J1748.9$-$2021, the blackbody has a radius of only $\sim$ 3 km and a temperature of $\sim$ 0.6 keV, while the temperature of the corona exceeds 14 keV \citep{2020MNRAS.492.4361S}.
   In the soft state of EXO 0748$-$676 and XTE J1701$-$462, this model constrains the spectral parameters well, the temperature of the corona being below 4.5 keV, and the temperature of blackbody below 0.5 keV \citep{2022MNRAS.510.4736K,2023MNRAS.525.4657J}.
   During the 2011 outburst of Aql X$-$1, as the source transitions from the hard state to the soft state, the temperature of the blackbody increases from $\sim$ 0.6 keV to $\sim$ 1.4 keV, and the electron temperature of the corona decreases from $\sim$ 22 keV to $\sim$ 3 keV \citep{2021JHEAp..31...12A}.
   Based on the fitting results of the sphere-corona model, as the sources transition from the hard state to the soft state, the blackbody temperature increases while the corona temperature decreases. 
 
   In MAXI J1807$+$132, both the disk-corona and sphere-corona models provide equally satisfactory fits to the data \citep{2025ApJ...978...12R}.
   The previous two examples of Aql X$-$1 also demonstrate the feasibility of both models.
   While both models effectively fit the continuum for specific spectral states, identifying the superior model based on its applicability across the entire outburst phase for various sources remains a challenging task.

   Aql X$-$1, first identified by \citet{1967Sci...156..374F}, is a transient NS-LMXB, classified as an atoll source \citep{1989A&A...225...79H,2000ApJ...530..916R}.
   It characteristically undergoes an outburst approximately once a year \citep{2017ApJ...848...13G}.
   This system has an orbital period of 18.9 hr \citep{1991A&A...251L..11C} and includes a K-type companion star \citep{1999A&A...347L..51C}.
   The estimated distance to the system is 6$\pm$2 kpc, and the orbital inclination is between 36$^\circ$ and 47$^\circ$ \citep{2017MNRAS.464L..41M}.
   A unique X-ray pulsation was detected at a frequency of 550.27 Hz \citep{2008ApJ...674L..41C}, occurring at the latter part of the rising phase of the 1998 outburst as observed by the Rossi X-ray Timing Explorer (\textit{RXTE}). 
   Additionally, during type-I X-ray bursts, burst oscillations were identified at around 549 Hz \citep{1998ApJ...495L...9Z,2008ApJS..179..360G}.
   By stacking the light curves of type-I X-ray bursts captured by \textit{RXTE/PCA} in this source, \citet{2013ApJ...777L...9C} discovered a deficit in the 30$-$50 keV band compared to the mean count rate before the burst, with a significance of 6 $\sigma$, and a delay the soft X-rays of 4.5$\pm$1.4 s.
   This deficit was interpreted as the corona being cooled down by the type-I X-ray bursts.

   The goal of our study is to identify the most appropriate model that describes the evolution of the spectral components throughout the outburst. 
   The frequent outbursts of Aql X$-$1 offer numerous opportunities to test more models.
   The 2023 outburst of Aql X$-$1 was caught by the Hard X-ray Modulation Telescope (\textit{Insight-HXMT}), presenting an unparalleled opportunity to fulfill our research aim.
   In this paper, we report the analysis of the 2023 outbust in Aql X$-$1.
   We describe the observations and data analysis in Section \ref{sec:Observation and data analysis}, and subsequently, we present our results in Section \ref{sec:Results}.
   In Section \ref{sec:Discussion and Conclusions}, we discuss and summarize our findings.

\section{Observation and data analysis}
\label{sec:Observation and data analysis}

   \begin{figure}
   \centering
   \includegraphics[width=\hsize]{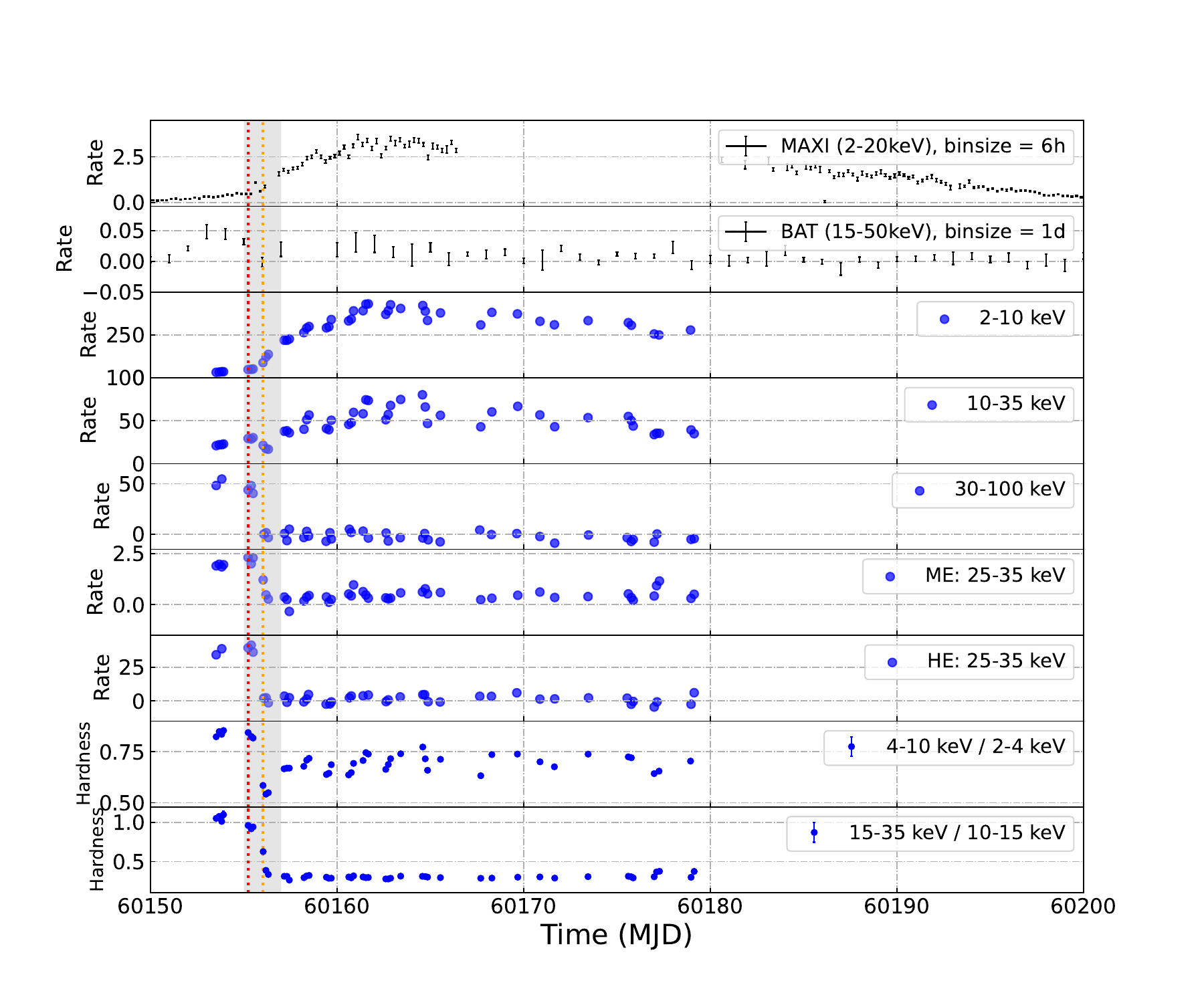}
      \caption{The light curves of the 2023 outburst of Aql X$-$1.
      From top to bottom, the first two panels show, respectively, the Monitor of All-sky X-ray Image (\textit{MAXI}) light curve at time bins of 6 hr and the Burst Alert Telescope (\textit{BAT}) light curve at time bins of 1 d.
      The following three panels show the light curves of LE (2$-$10 keV), ME (10$-$35 keV) and HE (30$-$100 keV) binned at one observation (1 ObsID, the average exposure time is around 2000 s), respectively.
      The next two panels show the 25$-$35 keV light curves of ME and HE binned at 1 ObsID.
      The last two panels show the hardness ratio curve using the 4$-$10 keV and 2$-$4 keV bands, and the 15$-$35 keV and 10$-$15 keV bands, respectively.
      The grey area of each panel roughly represents the duration of the state transition (MJD 60155$-$60157), which includes 6 observations.
      The red and orange dashed lines represent the burst \#1 and \#2, respectively.
      }
         \label{figure1}
   \end{figure}

   \begin{figure}
   \centering
   \includegraphics[width=\hsize]{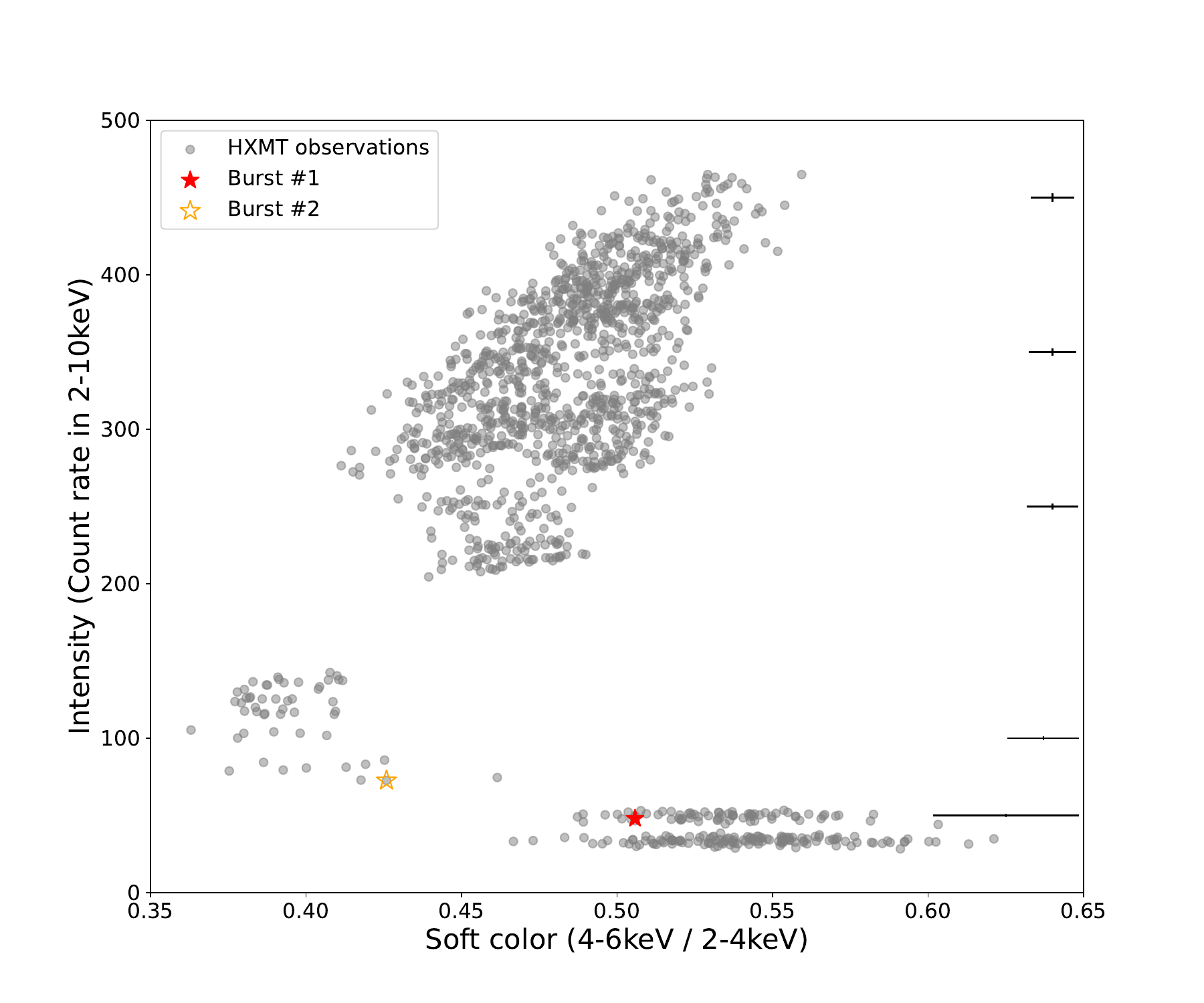}
      \caption{The hardness intensity diagram with time bins of 64 s from \textit{Insight-HXMT} observations of Aql X$-$1.
       The soft color is the ratio of the count rate in the 4$-$6 keV and 2$-$4 keV bands.
       The intensity is the count rate in the 2$-$10 keV.
       The red and orange stars represent the burst \#1 and \#2, respectively.
       The black crosses on the right side show the typical error bars at each intensity level.
              }
         \label{figure2}
   \end{figure}

   \begin{table}
        \bc
        \begin{minipage}[]{100mm}
	\caption[]{Overview of two type-I X-ray bursts.
		\label{table1}}\end{minipage}
   \setlength{\tabcolsep}{1pt}
   \small
    \begin{tabular}{ccccccccc}
     \hline\noalign{\smallskip}
             & & Onset & Rise & e-folding & & & & Persistent flux \\
             
             Burst & ObsID & date & time & time & Peak flux   & $\tau$ & $\Gamma$ & (2-100keV) \\
             
             & & (MJD) & (s) & (s) & ($\times10^{-8}$erg s$^{-1}$cm$^{-2}$) & (s) & & ($\times10^{-9}$erg s$^{-1}$cm$^{-2}$) \\
            \hline
            1 & P050426000201 & 60155.16570 & 5 & 32.52 & 2.93$\pm$0.61 & 27.37& 1.65$\pm$0.06 & 9.40$\pm$0.96  \\
            2$^{*}$ & P050426000301 & 60155.97533 & 10 & 6.51 & 2.90$\pm$0.59 & 13.33 & 2.00$\pm$0.05 & 7.86$\pm$0.53 \\
            \hline
	\end{tabular}
   \ec
   \tablecomments{0.86\textwidth}{The e-folding time, peak flux, and equivalent duration ($\tau$) were obtained by the standard method.
   The fitting model of the pre-burst spectrum is $\sc TBabs*powerlaw$.
   Notes: ${*}$ indicates the burst out of the GTIs.}
   \end{table}

   \textit{Insight-HXMT} \citep{Zhang2020} contains three payloads: the High Energy X-ray Telescope (HE, 20$-$250 keV), the Medium Energy X-ray Telescope (ME, 5$-$40 keV), and the Low Energy X-ray telescope (LE, 1$-$12 keV).
   We employed the \textit{Insight-HXMT} Data Analysis software (HXMTDAS) v2.06 for data reduction\footnote{http://hxmtweb.ihep.ac.cn/SoftDoc.jhtml}.
   The version of the calibration database (CALDB) is v2.07\footnote{http://hxmten.ihep.ac.cn/caldb/629.jhtml}.
   
   Aql X$-$1 started an outburst in July 2023 \citep{2023ATel16187....1A} that persisted for approximately 50 days. 
   From 2023 July 28 (MJD 60153) to 2023 August 23 (MJD 60179), a total of 439 ks of monitoring observations were obtained with \textit{Insight-HXMT}.
   These observations covered the onset and initial decline phases of this outburst, as shown by the LE, ME, and HE light curves in Figure~\ref{figure1}.
   To illustrate the trajectory of this outburst in the hardness intensity diagram (HID), we extracted 2$-$10, 2$-$4, and 4$-$6 keV light curves of these observations, each point with time bins of 64 s, as shown in Figure~\ref{figure2}.
   When analyzing observations containing bursts, we excluded the data from the burst part to generate the continuum spectra.
   The continuum spectra of LE and ME was re-binned using the the ftool $\sc grppha$ with a minimum of 1500 counts per grouped bin.
   For the HE spectra, we divided the energy channels into three ranges: below 25 keV, 25 to 50 keV, and above 50 keV.
   The channels were binned into 1 bin, 3 bins, and 11 bins, respectively, for each of these ranges.
   
   To search for type-I X-ray bursts, we extracted net light curves with a time bin of 1 s in the 1$-$10 keV and 10$-$35 keV bands for LE and ME, respectively. 
   During this outburst, we identified two type-I X-ray bursts occurring at the onset of the rising phase, with one falling outside the good time intervals (GTIs), as marked by the red and orange dashed lines in Figure~\ref{figure1}.
   To obtain the characteristics of the burst profile, we applied the method described by \cite{2024MNRAS.529.1585Y}.
   The onset time, the rise time, and the duration are summarized in Table~\ref{table1}.
   We also generated light curves with a bin size of 4 s in the 30$-$70 keV to investigate the variation in the hard X-ray band during the bursts.
   These light curves were subtracted by the 64-s mean count rate before the burst to avoid changes within a single observation.
   For the spectra before and during the burst, we employed the method in \citet{2024MNRAS.529.1585Y}.
   The LE and ME spectra were re-binned with a minimum of 20 counts per grouped bin using the same tool.

   Before performing the time-resolved spectral fitting, we added a systematic uncertainty of 1 percent, 2 percent, and 1 percent to the LE, ME, and HE spectra, respectively, based on the information provided by the \textit{Insight-HXMT} team \citep{2020JHEAp..27...64L}.
   XSPEC, V12.13\footnote{https://heasarc.gsfc.nasa.gov/docs/xanadu/xspec/index.html} was utilized for spectral analysis.
   We chose the energy bands 2$-$10 keV, 8$-$30 keV, and 25$-$100 keV for the LE, ME, and HE instruments, respectively.
   For the observations without emission above 25 keV, we only used the LE and ME data.
   The errors of the parameter values were calculated at the $1\sigma$ confidence level.
   In the spectral fitting, we used the $\sc TBabs$ model with the photoionization cross-sections of \citet{Verner1996}, and the solar abundances \citep{Wilms2000} to account for the interstellar absorption.
   Since our data cannot effectively constrain the hydrogen column density well, we fixed it at 0.36 $\times 10^{-22}{\rm cm}^{-2}$ \citep{2012PASJ...64...72S}.

   \begin{table}
   \bc
   \begin{minipage}[]{100mm}
   \caption[]{The fitting models in this work.\label{table2}}\end{minipage}
   \setlength{\tabcolsep}{1pt}
   \small
    \begin{tabular}{cc}
     \hline\noalign{\smallskip}
   Model 1	&  $\sc TBabs*(Thcomp*diskbb+bbodyrad)$ \\
   Model 2	&  $\sc TBabs*(Thcomp*bbodyrad+diskbb)$ \\
   Model 3	&  $\sc TBabs*(bbodyrad+diskbb+bbodyrad)$ \\
   \noalign{\smallskip}\hline
   \end{tabular}
   \ec
   \tablecomments{0.86\textwidth}{We employed the models 1 and 2 to fit the spectra of Aql X$-$1 before the transitional state, and Model 3 to fit the spectra in the soft state.}
   \end{table}

   To investigate the evolution of the corona, the accretion disk, and other potential components (such as the BL and the hot spot) throughout the outburst in Aql X$-$1, the majority of studies employed two types of model: one incorporates a Comptonized disk blackbody and a blackbody \citep{2012PASJ...64...72S}, while another one assumes the blackbody emission is Comptonized by the corona, with the disk blackbody making a direct contribution to the observed emission \citep{2012PASJ...64...72S,2017PASJ...69...23O,2020JHEAp..25...10G,2023MNRAS.521.4490F}.
   In our study, we adopted the $\sc Thcomp$ component for a precise representation of the thermal Comptonization spectra, as validated by \citet{2020MNRAS.492.5234Z}.
   We employed the first type model (disk-corona model) and the second type model (sphere-corona model) for testing.
   The detailed configurations of these models, designated as Model 1 and Model 2, are presented in Table~\ref{table2}.
   The corona was described using $\sc Thcomp$, $\sc diskbb$ represented the accretion disk, and $\sc bbodyrad$ represented the blackbody around/from the NS surface.
      
   At the beginning of the spectral testing, we applied these models to fit all observations, and found that they both fitted the data well before the source entered the soft state (MJD 60157).
   In the soft state, we opted to proceed with Model 2 for the following tests.
   The reason for utilizing Model 2 is explained in Section \ref{sec:Appropriate model for the 2023 outburst}.
   In the test, we permitted all parameters of Model 2 to vary freely.
   However, with the disappearance of the hard X-ray emission above 25 keV (MJD 60156), this approach failed.
   Consequently, we attempted to fix some parameters of the Compton component, which might roughly remain constant.
   
   We conducted two trials to investigate the evolution of the Compton component in Model 2.
   In the first trial, we fixed the covering fraction (Cov\_frac) during the soft state.
   Based on the fitting results before the soft state, the covering fraction decreased from 100\% to around 12\%.
   We then set the covering fraction at 12\% to fit the subsequent observations.
   In the fitting results, the temperature of the corona stabilized at 2$-$3 keV.
   Additionally, the photon index varies from 1 to 2.
   However, this range is significantly lower than the expected values for the soft state case \citep{2004ApJ...608..444M,2018A&A...616A..23D,2023MNRAS.521.4490F}.
   In the second trial, we fixed the temperature of the corona (kT$_{\rm e}$) during the soft state.
   To estimate the possible upper limit of the corona temperature, we combined two neighbouring  observations (P050426000302 and P050426000303) before the soft state to enhance the statistical accuracy.
   These observations exhibited similar count rates and hardness ratios across different energy bands, as shown in Figure~\ref{figure1}.
   The best-fitting parameters of the corona were as follows: kT$_{\rm e}$ = 6.18 $\pm$ 2.43 keV and Cov\_frac = 0.87 $\pm$ 0.13.
   Considering this possible upper limit of the corona temperature, we tentatively selected a range of fixed kT$_{\rm e}$ values from 3 keV to 5 keV, with a step size of 0.5 keV.
   In the fitting results, as kT$_{\rm e}$ increases, $\Gamma$ also increases, with the values of kT$_{\rm e}$ at 4.5 keV and 5 keV nearly overlapping.
   The covering fraction exhibits a random behavior at lower kT$_{\rm e}$ values (3$-$4 keV), but stabilizes around 1 at higher kT$_{\rm e}$.
   The parameters of the blackbody and disk blackbody components remain consistent across different corona temperatures.
   These trials, along with the absence of photons above 25 keV (as illustrated in the fifth, sixth and seventh panels of Figure~\ref{figure1}), indicate that the hard emission is missing during the soft state, suggesting that the Compton component is not needed in this state.

   Subsequently, we attempted to fit the continuum after MJD 60156 by removing the Compton component, but the model did not fit the data well.
   The data-to-model ratio showed a significant deviation below 3 keV, which indicates the need for an additional thermal component.
   We tentatively introduced an additional blackbody component to formulate Model 3 ($\sc TBabs*(bbodyrad+diskbb+bbodyrad)$).
   This model comprises two blackbodies (the high-temperature blackbody is bb$_1$, and the low-temperature blackbody is bb$_2$) and a disk blackbody.
   This model described the continuum well.
   We adopted a distance of 5.2 kpc and an inclination angle of 45$^\circ$ for the disk to calculate the radius of the blackbody and the inner disk. 
   Utilizing the formulas provided in the manual of XSPEC\footnote{https://heasarc.gsfc.nasa.gov/docs/xanadu/xspec/index.html} for the $\sc bbodyrad$ and $\sc diskbb$ models, we converted the normalization parameter to the radius in kilometers.
   We utilized the $\sc cflux$ component to calculate the unabsorbed 2$-$100 keV of the total flux of the model, as well as the flux of each component within the model.

   For the spectral analysis during the type-I X-ray burst phase, we applied the same methods of \citet{2024MNRAS.529.1585Y}.
   The overall information of the two type-I X-ray bursts is shown in Table~\ref{table1}.
   To calculate the significance of the hard X-ray (30$-$70 keV) deficit during the type-I X-ray bursts, we selected the deficit interval during the burst, as shown in the gray area in Figure~\ref{figure1}.
   Its significance is the ratio of the deficit area and the total error following propagation \citep{2018ApJ...864L..30C}.
   We used $\sc crosscor$ \footnote{http://heasarc.gsfc.nasa.gov/ftools/fhelp/crosscor.txt} to calculate the cross-correlation between the LE light curve (1$-$10 keV, 1$-$s resolution) and the HE light curve (30$-$70 keV, 1$-$s resolution).
   To estimate the time lag and its error, we used the GAUS model in QDP \footnote{https://heasarc.gsfc.nasa.gov/ftools/others/qdp/qdp.html} to fit the distribution of the time delay.

\section{Results}
\label{sec:Results}

\subsection{Light curves and hardness-intensity diagram}

   In the first two panels of Figure~\ref{figure1}, we used the light curves of \textit{MAXI} and \textit{Swift-BAT} to illustrate the entire phase of the 2023 outburst in Aql X$-$1.
   The next three panels show the light curves of the LE (2$-$10 keV), ME (10$-$35 keV), and HE (30$-$100 keV) instruments on board \textit{Insight-HXMT}.
   As the LE and ME light curves increase simultaneously, the HE light curve shows a clear drop from MJD 60156 and remains at around zero after that.
   This rapid decrease in the hard X-ray emission during the early phase of this outburst has also been observed in previous outbursts of Aql X$-$1 \citep{2007ApJ...667.1043Y,2023MNRAS.521.4490F}.
   This similar behavior also occurred in other sources, such as IGR J17473$-$2721 \citep{2010A&A...510A..81C} and 4U 1608$-$52 \citep{2012PASJ...64..128A}.

   The last two panels of Figure~\ref{figure1} represent the hardness ratio for the LE (4$-$10 keV / 2$-$4 keV) and ME (15$-$35 keV / 10$-$15 keV) instruments, respectively.
   As the outburst progresses, the LE hardness first shows a dip from MJD 60156 and then slowly decreases, while the ME hardness rapidly decreases and then remains at a low level. 
   During this dip stage (the grey area in Figure~\ref{figure1}), the 30$-$100 keV light curve rapidly decreases to zero.
   So, this stage represents the state transition from the hard to the soft state. 
   These phenomena also were observed in the 2009 outburst of Aql X$-$1 \citep{2010ApJ...716L.109M}.
   
   The spectral state evolution of Aql X$-$1 can be intuitively distinguished in the HID, as shown in Figure~\ref{figure2}.
   The source evolves from the bottom right corner to the left and then to the upper right corner in the plot.
   This trace is similar to the HID reported by \cite{2022MNRAS.510.1577G}.
   They defined the intensity as the count rate in the 0.5$-$10.0 keV band and the hardness ratio as the count rate ratio between the 3.8$-$6.8 keV and 2.0$-$3.8 keV bands. 
   The observations in our trace cover the low/hard state (before MJD 60155), the state transition (between MJD 60155 and MJD 60157), and the high/soft state (after MJD 60157) of Aql X$-$1.
   The two bursts took place in the transition state.

\subsection{Broad-band spectral analysis}

   \begin{figure}
   \centering
   \includegraphics[width=\hsize]{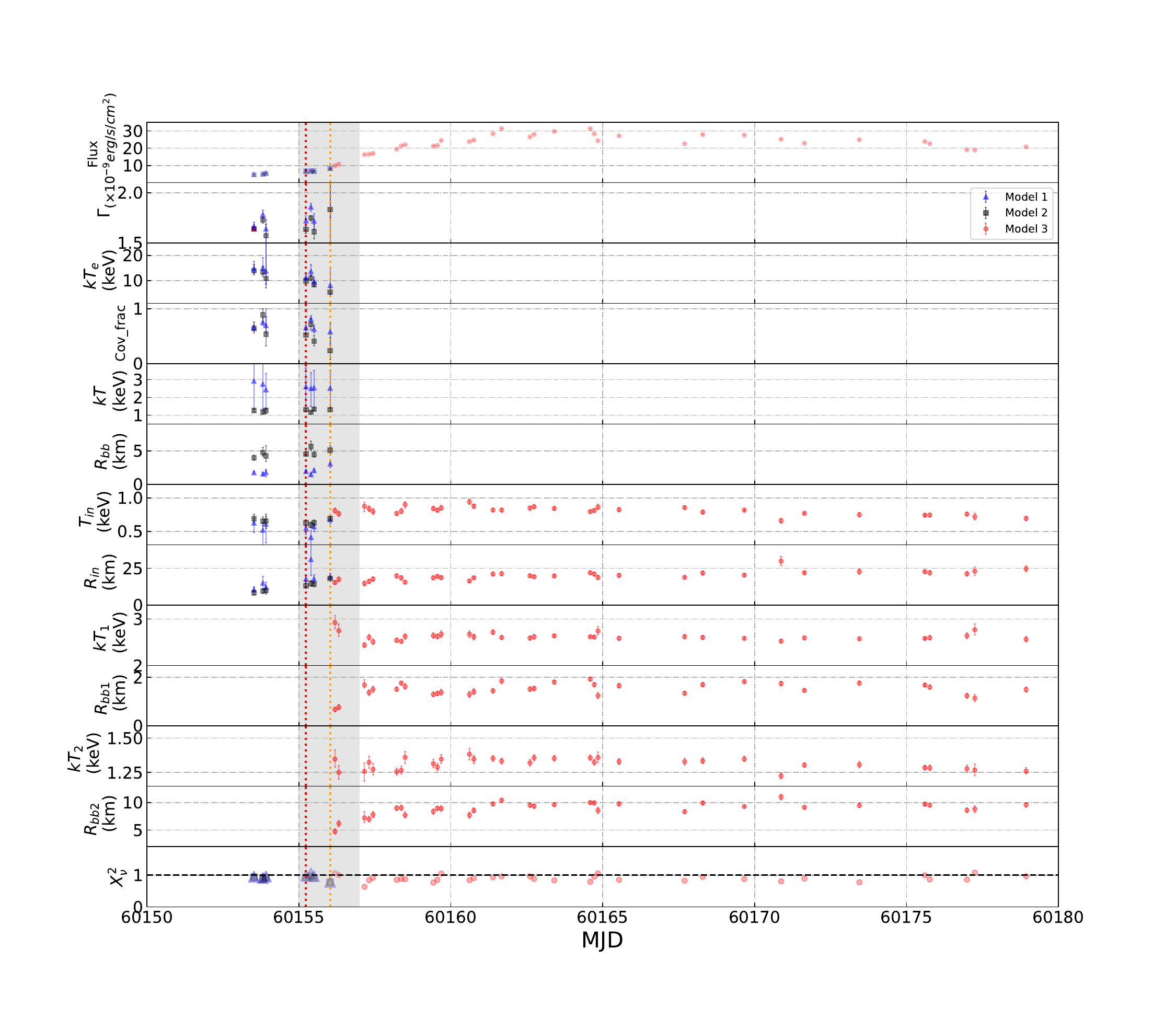}
      \caption{The fitting results of three models in Aql X$-$1.
      From top to bottom, each panel represents the unabsorbed flux ($\times10^{-9}$erg s$^{-1}$cm$^{-2}$), the photon index ($\Gamma$), the electron temperature (kT$_{\rm e}$), the covering fraction of the corona (Cov\_frac), the temperature (kT) and the radius (R$_{\rm bb}$) of the blackbody, the temperature at the inner disk radius (T$_{\rm in}$) and the inner disk radius (R$_{\rm in}$), the temperature (kT$_{1}$) and radius (R$_{\rm {bb1}}$) of the high-temperature blackbody, and the temperature (kT$_{2}$) and radius (R$_{\rm {bb2}}$) of the low-temperature blackbody.
      The last panel is the reduced chi-square ($\chi_v^2$).
      The blue triangles, black squares, and red points represent the fitting results of Model 1, Model 2, and Model 3, respectively.
      The grey area of each panel roughly represents the duration of the state transition.
      The red and orange dashed lines represent the time of the burst \#1 and \#2, respectively.
              }
         \label{figure3}
   \end{figure}

   In first eight panels of Figure~\ref{figure3}, we plotted the fitting results of Model 1 (blue triangles) and Model 2 (black squares) before the soft state (MJD 60157) in the 2023 outburst.
   In the second panel, the value of the photon index ($\Gamma$) in both Model 1 and Model 2 gradually increased.
   The temperature and covering fraction of the corona exhibited decreasing trends in the subsequent two panels.
   The blackbody temperature for Model 1 was around 2.5 keV, notably higher than the value (around 1.2 keV) measured in Model 2.
   The blackbody radius in both models were around 2 km and 5 km, both of which are below the typical radius of a NS.
   This indicated that the emission of the blackbody may come from parts of the surface of the NS.
   These two models both fitted the continuum well during the hard state.
   
   After the disappearance of the hard emission, we employed Model 3; the fitting results are depicted by the red points in Figure~\ref{figure3}.
   The trend of the parameters of the disk from Model 3 are consistent with the trends observed in the previous two models, without any abrupt shifts.
   For Model 3, the temperature of the inner disk gradually decreases as the outburst evolves, while the radius of the inner disk expands to 25 km.
   Concurrently, the high-temperature blackbody shows a more or less consistent temperature at approximately 2.5 keV and a radius of approximately 1.8 km. 
   The constancy of this component implies that the thermal emission may originate from a localized area on the surface of the NS, which remains invariant throughout the soft state.
   Simultaneously, the temperature evolution of the low-temperature blackbody exhibits an initial rise to approximately 1.4 keV, followed by a gradual decline to approximately 1.25 keV. 
   Correspondingly, the radius of this blackbody undergoes an expansion from approximately 7.5 km to 10 km.
   This indicates that the increasing radius of the low-temperature blackbody contributes more to the changes of the total flux than the high-temperature blackbody during the outburst.

   \begin{figure}
   \centering
   \includegraphics[width=12.0cm, angle=0]{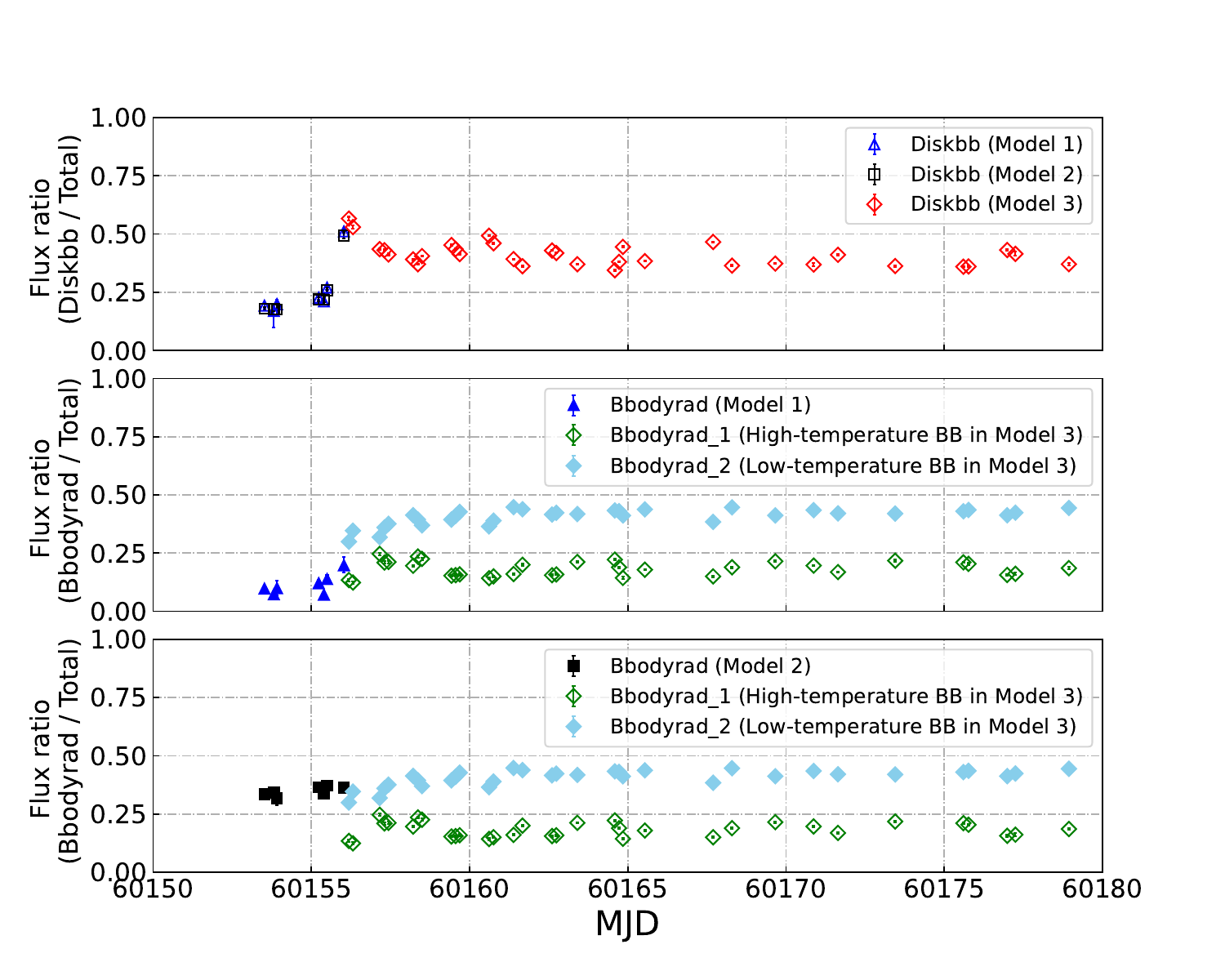}
      \caption{The evolution of flux ratio of the $\sc diskbb$ and $\sc bbodyrad$ for the three models.
      The flux ratio is the ratio of the unabsorbed flux of the given component to the unabsorbed total flux.
      In the upper panel, the blue triangles, black squares, and red diamonds represent the flux ratio of the $\sc diskbb$ in the three models.
      In the next two panels, the blue triangles, black squares, and diamonds represent the flux ratio of the $\sc bbodyrad$, while the green and skyblue diamonds represent the high-temperature blackbody  (Bbodyrad\_1) and the low-temperature blackbody (Bbodyrad\_2) in Model 3, respectively.
      }
         \label{figure4}
   \end{figure}

   We calculated the ratio of the unabsorbed flux from each component to the total unabsorbed flux for each model, as shown in Figure~\ref{figure4}.
   This plot only presents data from three models in distinct parts, paralleling the results illustrated in Figure~\ref{figure3}.
   In the upper panel, the flux ratios for the disk component in Model 1 and Model 2 exhibit a similar behavior.
   Across the three models, this ratio undergoes a rapid rise from approximately 0.2 ($\sim$ MJD 60155) to nearly 0.6 ($\sim$ MJD 60156), eventually settling around 0.4 (after $\sim$ MJD 60157).
   This trend suggests that the emission of the disk undergoes significant fluctuations primarily during the transitional state.
   In the second panel, the flux ratio attributed to the blackbody component in Model 1 increases from 0.1 to 0.2 during the transitional state. 
   Conversely, in Model 2, this ratio (in the last panel) almost stabilizes below 0.4.
   For Model 3, both the high-temperature and low-temperature blackbody components demonstrate a steady rise in their respective flux ratios. 
   The flux ratio of the high-temperature component plateaus at 0.2, while the low-temperature component stabilizes at 0.4.

\subsection{Hard X-ray deficit in the Type-I X-ray bursts}

   \begin{figure}
       \begin{minipage}{0.52\linewidth}
        	\includegraphics[width=0.9\columnwidth]{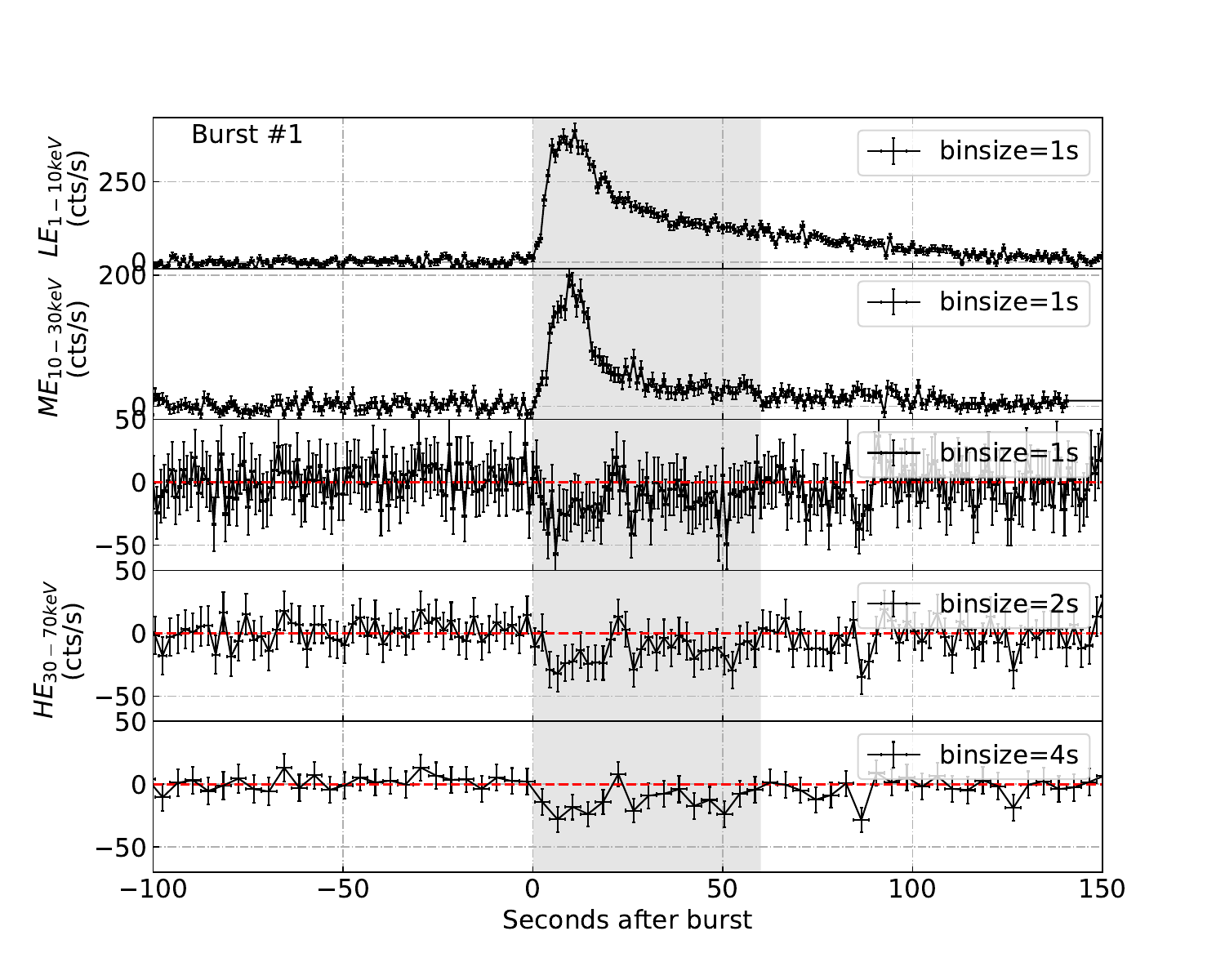}
       \end{minipage}
       \quad
       \begin{minipage}{0.52\linewidth}
   	\includegraphics[width=0.9\columnwidth]{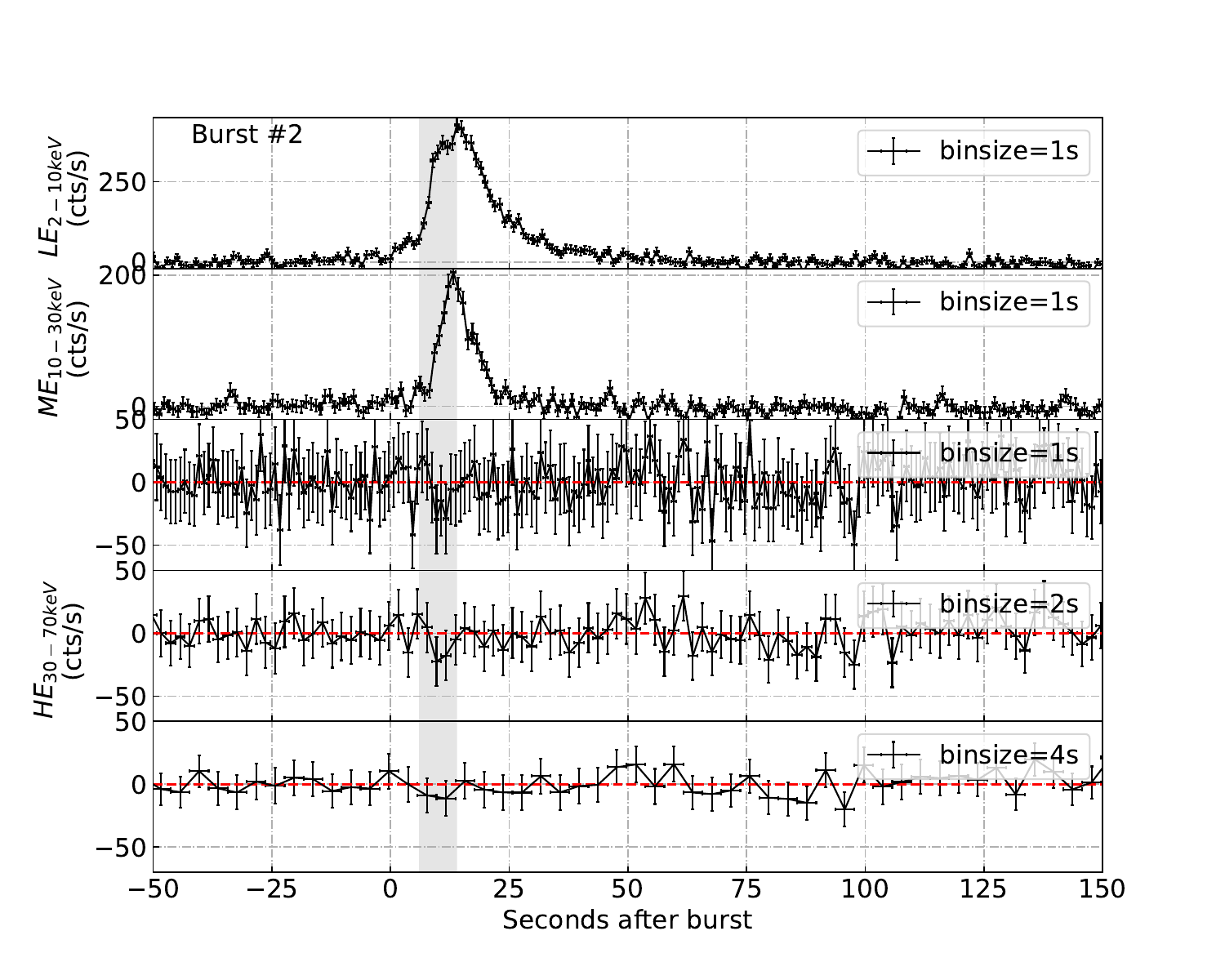}
       \end{minipage}
       \caption{The light curves of burst \#1 and burst \#2 in Aql X$-$1 with \textit{Insight-HXMT}.
       The first two panels in the left plot show the LE (1$-$10 keV) and ME (10$-$30 keV) light curves.
       The following three panels show HE (30$-$70 keV) light curves with different binsize (1s, 2s, 4s).
       To produce these light curves we subtracted the mean count rate of the pre-burst part and corrected by the effect of dead time.
       }
   \label{figure5}
   \end{figure}

   In Figure~\ref{figure5}, we present the light curves, with 1-s resolution (2-s and 4-s resolution for HE), of two type-I X-ray bursts as observed by three instruments in \textit{Insight-HXMT}.
   Burst \#1 lasts approximately 100 s in the 1$-$10 keV band and shows a significant deficit in the 30$-$70 keV.
   The observed deficit with a duration of 60 s (the grey shaded region of the left panel in Figure~\ref{figure5}) in burst \#1 reaches a significance of 4.82 $\sigma$.
   Furthermore, we measured a time delay of ${-0.91}_{-1.01}^{+0.91}$ s between the soft and hard emissions.
   If we take into account the error bar, this value could be zero or negative, which suggests that the time delay in this burst is not obvious.
   The burst \#2 has similar peak count rate, a shorter duration, and a longer rise time in the 1$-$10 keV light curve.
   This burst exhibits a deficit lasting for 8 s (indicated by the grey shaded region in the right panel of Figure~\ref{figure5}), with a significance of 1.01 $\sigma$. 
   Given the low statistical significance of this feature, we did not perform a correlation analysis.

\section{Discussion and Conclusions}
\label{sec:Discussion and Conclusions}

   We analyzed the 2023 outburst of the transient NS-LMXB Aql X$-$1 using observations from \textit{Insight-HXMT}, covering its evolution from the hard state through the transition state and into the soft state. 
   To model the continuum before the soft state, we employed two approaches, both revealing a decline in electron temperature and covering fraction of the Comptonized component. 
   In the soft state, the double blackbody model provided an excellent fit to the observed continuum. 
   Additionally, we detected two bursts before the transition to the soft state, with burst \#1 exhibiting a pronounced deficit in hard X-rays compared to burst \#2.

\subsection{Appropriate model for the 2023 outburst}
\label{sec:Appropriate model for the 2023 outburst}

   We utilized two models to describe the behavior of the continuum in the 2023 outburst.
   The first model comprises a Comptonized disk and a blackbody (Model 1), while the second includes a disk and a Comptonized blackbody (Model 2).
   Both models effectively capture the hard state, but they fail to constrain the Compton component once the hard emission disappears during the transition state. 
   In the soft state, we found the double blackbody model (Model 3) successfully fitted the spectra.
   This model consists of a low-temperature blackbody, a high-temperature blackbody and a disk.

   Given that our data cover the phase from the rise to the initial decline of the outburst in Aql X$-$1, we have collected several previous studies that describe similar outburst phases for comparison.
   The 2000 outburst, observed by Rossi X-Ray Timing Explorer/Proportional Counter Array (\textit{RXTE/PCA}, 2$-$60 keV), was analyzed by \citet{2004ApJ...608..444M} using model $\sc wabs*(diskbb+powerlaw)$.
   They interpreted the power-law component as emission from the corona.
   As the source transitioned from the hard state to the soft state, both the temperature (from $\sim$ 0.5 keV to $\sim$ 2 keV ) and the radius of the inner disk increased.
   These trends align with our results, though the inner disk temperature in our work only increases from 0.5 keV to 1 keV.
   Recent analysis of the 2019 and 2020 outbursts by \citet{2024MNRAS.532.3961P} employing  the disk-corona model ($\sc TBabs*(gauss+gauss+thcomp*diskbb)$) on Neutron Star Interior Composition Explorer (\textit{NICER}, 0.5$-$10 keV) data revealed similar behavior: the inner disk temperature rose from $\sim$ 0.6 keV to $\sim$ 0.9 keV while the radius expanded from $\sim$ 14 km to $\sim$ 20 km (the covering fraction is fixed at 0.5). 
   Though consistent with our observed trends, our work shows a larger maximum disk radius ($\sim$ 25 km).
   However, alternative modeling approach (the sphere-corona model) yield conflicting results. 
   \citet{2023MNRAS.521.4490F} analyzed the 2009 outburst using joint \textit{SWIFT/XRT} (0.5$-$10 keV) and \textit{RXTE/PCA} (2.6$-$23 keV) observations with the model $\sc TBabs*(nthcomp+diskbb)$.
   They reported a decreasing inner disk radius (from $\sim$ 22 km to $\sim$ 10 km) despite comparable temperature evolution (from $\sim$ 0.4 keV to $\sim$ 0.9 keV), which contrasts with our findings.
   Similar discrepancies are also apparent in the analysis of the 2011 outburst by  \citet{2017PASJ...69...23O}, who utilized \textit{Suzaku} (0.8$-$60 keV) observations and fitted the spectra with $\sc wabs*(nthcomp+diskbb)$.
   In this analysis, the radius decreased from $\sim$ 35 km to $\sim$ 20 km while temperatures increased from $\sim$ 1 keV to $\sim$ 1.5 keV.
   Our work differs from these earlier studies by proposing that the Compton component vanishes following the disappearance of hard emission, and we employ a double blackbody model to describe the continuum. 
   
   We observed that as the source transitions from the hard state to the soft state, the inner disk radius gradually increases. 
   This finding is not only inconsistent with some previous studies but also deviates from the predictions of the standard disk theory. 
   To further investigate this phenomenon, we conducted several tests on the inner disk radius under different spectral states.
   In the hard state, we used Model 2 to fit the spectrum of the first observation (P050426000101), obtaining a disk normalization of $\sim$ 181. 
   In contrast, during the soft state, we fitted the spectrum of the observation (P050426001701) with Model 3, resulting in a disk normalization of $\sim$ 2343. 
   The disk normalization in the soft state is significantly higher than that in the hard state.
   After removing the additional blackbody component from Model 3 and refitting the same spectrum in the soft state, the disk normalization was measured at $\sim$ 109. 
   This value is lower than that in the hard state, suggesting that when the source transitions from the hard state to the soft state, the disk moves closer to the NS.
   This trend aligns with the standard disk theory.
   Additionally, this value is lower than that obtained with Model 3, prompting further investigation.
   The fitting results show a chi-squared/d.o.f. of 302.08/252 for the Model 3 without the additional blackbody component, compared to 201.69/250 for Model 3.
   We employed the  $\sc ftest$ to calculate the F-statistic and its probability, resulting in  an F-statistic value of 62.22 and a probability of 1.18 $\times$ 10$^{-22}$. 
   This outcome indicates that the inclusion of the additional blackbody component significantly improves the fit.
   Consequently, the presence of the additional blackbody component affects the measurement of the inner disk radius, suggesting that the trend we observe may not accurately reflect the actual changes in the inner disk radius.
   Furthermore, the energy range we used, starting at 2 keV, may impact the measurement of the inner disk radius.

   Building on previous analyses, we explore the appropriate model through an evaluation of the correspondence between the blackbodies in the first two models and the blackbodies in Model 3, as well as whether the appearance of the BL and the hot spot in Model 3 is reasonable.
   As previous spectral studies of Aql X$-$1 have shown, the temperature of the BL typically ranges from 1.6 keV to 3 keV and the temperature of the hot spot could exceed 3 keV \citep{2017ApJ...848...13G,2023MNRAS.521.4490F}.
   In other NS-LMXB sources, such as 4U 1608$-$52, 1A 1744$-$361 and Cir X$-$1, the temperature of the BL was below 2 keV \citep{2017MNRAS.467..290A,2024ApJ...966..232N,2024ApJ...961L...8R}.
   From the above, we conclude that the temperature of the hot spot is higher than part of the BL.
   The origin of the high-temperature blackbody emissions could be attributed to the hot spot, and the low-temperature blackbody could correspond to the BL.
   Next, we use two assumptions to discuss the potential scenarios that may occur during the state transition.

   Firstly, assuming that Model 1 (the disk-corona model) accurately represents the scenario in the hard state.
   Our results show that the fraction of the disk in Model 1 is $\sim$ 20 percent, and the ratio of the blakbody flux to the disk flux is $\sim$ 50 percent.
   During the same state in XTE J1710$-$281, \citet{2020MNRAS.496..197S} measured these values at $\sim$ 22 percent and $\sim$ 1 percent.
   Only the ratio of the blackbody flux to the disk flux differs significantly from ours.
   During the transient state, the temperature of the blackbody in Model 1 remains constant at $\sim$ 2.5 keV, and the high-temperature blackbody in Model 3 starts at $\sim$ 3 keV.
   At the same time, the fraction of the blackbody in Model 1 increases to near the beginning of the fraction of the high-temperature blackbody in Model 3.
   If we consider the temperature and the flux ratio of blackbodies to be continuously evolving, then the blackbody in Model 1 corresponds the high-temperature blackbody in Model 3.
   We infer that the high-temperature blackbody corresponds to the hot spot, while the low-temperature blackbody would be indicative of the BL.
   As the source enters the soft state, the BL suddenly appears at $\sim$ 30 percent of the total flux, and increases to the level of the disk ($\sim$ 40 percent).

   Secondly, we consider Model 2 as the appropriate representation of the hard state.
   In the hard state, the fraction of the disk in Model 2 is $\sim$ 20 percent, which means that the fraction of the Comptonized blackbody is $\sim$ 80 percent.
   It shows that the Comptonized blackbody dominates the emission in the hard state.
   During the transition state, the temperature of the blackbody in Model 2 remains at $\sim$ 1.2 keV, and the low-temperature blackbody in Model 3 starts at $\sim$ 1.3 keV.
   Taking into account the continuous evolution, the blackbody in Model 2 corresponds to the low-temperature blackbody in Model 3.
   According to the correspondence between spectral components mentioned earlier, when the source enters the soft state, the contribution of the BL increases from $\sim$ 35 percent to $\sim$ 40 percent of the total flux. 
   The hot spot suddenly appears at $\sim$ 10 percent of the total flux.
   
   Based on the existing flux ratio relation, it is difficult to clearly determine which assumption is better.
   Transient unique X-ray pulsation were observed during the soft state (around MJD 50882) of the 1998 outburst in Aql X$-$1 \citep{2008ApJ...674L..41C}; this signal may be attributed to the appearance of a hot spot.
   In Aql X$-$1, the oscillations produced by the hot spot only occurred during the soft state \citep{2004ApJ...608..930M}.
   In XTE J1701$-$462 and 4U 1702$-$429, the BL could exist across all spectral state \citep{2009ApJ...696.1257L,2024MNRAS.529.4311B}, and the emission of the BL will be enhanced by the increasing accretion rate \citep{2014arXiv1412.1164D,2020A&A...638A.142A}.
   Therefore, in this assumption, the appearance of hot spot in the soft state seems more reasonable.
   Consequently, Model 2 would be more suitable than Model 1 for describing the hard state.

   Based on the fitting results of Model 2 and Model 3, when Aql X$-$1 transitions from the hard to the soft state in the 2023 outburst, the changes in the accretion environment surrounding the neutron star are characterized by the following: the inner edge of the accretion disk progressively retreats from the vicinity of the neutron star, its temperature steadily increases, and emission strengthens.
   The corona shows a decrease in the temperature and covering fraction.
   The temperature and radius of the BL exhibit a gradual increase, and similarly, the radius of the hot spot also increases.

\subsection{The corona cooling by the type-I X-ray burst}

   Our analysis reveals two type-I X-ray bursts in Aql X$-$1, both exhibiting comparable peak fluxes (Table~\ref{table1}).
   Notably, we detected burst \#1 during the hard state, which demonstrates a significant hard X-ray deficit of $\sim$ 4.82 $\sigma$.
   In the transition state, we observed burst \#2, which shows no significant deficit ($\sim$ 1.01 $\sigma$). 
   This finding aligns with the 6 $\sigma$ deficit reported by \citet{2013ApJ...777L...9C} through stacking 26 bursts observed in the hard state by \textit{RXTE} prior to 2012.
   The consistency between their detection of the hard X-ray deficit and our results provides strong confirmation of the essential association between the hard X-ray deficit and bursts in the hard state.
   Particularly noteworthy is the single-burst detection capability demonstrated by the HE. 
   With its large effective area, HE enables unprecedented tracking of the evolution of the corona in real time during  one single burst, such as \citet{2018ApJ...864L..30C}.

   For burst \#1, spectral analysis revealed the following parameters: a blackbody temperature of the burst peak of 2.26$\pm$0.10 keV, a temperature of the corona of 11.51$\pm$1.67 keV and a covering fraction of 0.71$\pm$0.07.
   For burst \#2, the corresponding values are 2.16$\pm$0.09, 6.00$\pm$2.25 keV and 0.33$\pm$0.13.
   The blackbody temperatures at the burst peak, which correspond to the temperature of the soft photons, show consistency between the two bursts.
   When the corona temperature drops from $\sim$ 11.51 keV to $\sim$ 6.00 keV, the thermal gradient between the corona and soft photon also decreases from $\sim$ 9.30 keV to $\sim$ 3.79 keV.
   Meanwhile, the covering fraction decreases from $\sim$ 0.71 to $\sim$ 0.33.
   The significance of the deficit also decreases from $\sim$ 4.82 $\sigma$ to $\sim$ 1.01 $\sigma$.
   The changes in these values indicate that a corona with a lower covering fraction and lower temperature cannot be effectively cooled by the burst.
   
   During burst \#1 in Aql X$-$1, we measured the time delay between the LE and HE light curves to be ${-0.91}_{-1.01}^{+0.91}$ s.
   In a previous study of the same source, \citet{2013ApJ...777L...9C} stacked 26 bursts and reported a time delay of 1.8$\pm$1.5 s.
   This value is inconsistent with our measurement.
   In another NS-LMXB, 4U 1636$-$53, \citet{2013MNRAS.432.2773J} stacked 36 bursts observed by \textit{RXTE} and measured a time delay of 2.4$\pm$1.5 s.
   However, \citet{2018ApJ...864L..30C} reported different value of 1.6$\pm$1.2 s for a single burst observed by \textit{Insight-HXMT}.
   These discrepancies show that the time delays measured from stacked bursts may differ from those measured from individual bursts, indicating that the averaging effect of stacking could influence the observed time delay values.
   In previous studies of other sources, such as IGR J17473$-$2721, KS 1731$-$260 and GS 1826$-$268, the time delays between soft and hard X-ray emissions were found to be 0.7$\pm$0.5 s, 0.9$\pm$ 2.1 s and 3.6$\pm$1.2 s \citet{2012ApJ...752L..34C,2014A&A...564A..20J,2014ApJ...782...40J}, respectively.
   The differences in these time delays suggest that differences in accretion states and corona geometries among these sources may contribute to the observed discrepancies in the time delay \citet{2018SSRv..214...15D}.

\normalem
\begin{acknowledgements}
   J. M is supposed by the National Key R\&D Program of China (2023YFE0101200), Yunnan Revitalization Talent Support Program (YunLing Scholar Award), and the National Natural Science Foundation of China (NSFC) grant 12393813. 
   GB acknowledge the science research grants from the China Manned Space Project.
   Lyu is supported by Hunan Education Department Foundation (grant No. 21A0096). 
   This work is supported by the National Key R\&D Program of China (2021YFA0718500).

\end{acknowledgements}

\bibliographystyle{raa}
\bibliography{bibtex}

\end{document}